\documentstyle[12pt]{article}

\font\twlgot =eufm10 scaled \magstep1
\font\egtgot =eufm8
\font\sevgot =eufm7

\font\twlmsb =msbm10 scaled \magstep1
\font\egtmsb =msbm8
\font\sevmsb =msbm7

\newfam\gotfam
\def\pgot{\fam\gotfam\twlgot}
\textfont\gotfam\twlgot
\scriptfont\gotfam\egtgot
\scriptscriptfont\gotfam\sevgot
\def\got{\protect\pgot}

\newfam\msbfam
\textfont\msbfam\twlmsb
\scriptfont\msbfam\egtmsb
\scriptscriptfont\msbfam\sevmsb

\def\pBbb{\relax\ifmmode\expandafter\Bb\else\typeout{You cann't use
Bbb in text mode}\fi}
\def\Bb #1{{\fam\msbfam\relax#1}}

\def\thebibliography#1{\bigskip\section*{\large
\bf References\\}\list
  {[\arabic{enumi}]}{\settowidth\labelwidth{#1}\leftmargin\labelwidth
    \advance\leftmargin\labelsep
    \usecounter{enumi}}
    \def\newblock{\hskip .11em plus .33em minus .07em}
    \sloppy\clubpenalty4000\widowpenalty4000
    \sfcode`\.=1000\relax}

\def\op#1{\mathop{{\it\fam0} #1}\limits}

\newcommand{\beq}{\begin{equation}}
\newcommand{\eeq}{\end{equation}}
\newcommand{\ben}{\begin{eqnarray}}
\newcommand{\een}{\end{eqnarray}}
\newcommand{\be}{\begin{eqnarray*}}
\newcommand{\ee}{\end{eqnarray*}}
\newcommand{\bea}{\begin{eqalph}}
\newcommand{\eea}{\end{eqalph}}

\newcommand{\cG}{{\got g}}

\newcommand{\gj}{{\got J}}

\newcommand{\gS}{{\got S}}

\newcommand{\gF}{{\got F}}

\newcommand{\gP}{{\got P}}

\newcommand{\cP}{{\cal P}}

\newcommand{\cL}{{\cal L}}

\newcommand{\cF}{{\cal F}}

\newcommand{\cS}{{\cal S}}

\newcommand{\bL}{{\bf L}}

\newcommand{\al}{\alpha}
\newcommand{\bt}{\beta}
\newcommand{\dl}{\delta}
\newcommand{\la}{\lambda}

\newcommand{\f}{\phi}
\newcommand{\vf}{\varphi}

\newcommand{\om}{\omega}

\newcommand{\m}{\mu}

\newcommand{\g}{\gamma}

\newcommand{\e}{\epsilon}
\newcommand{\ve}{\varepsilon}

\newcommand{\si}{\sigma}

\newcommand{\w}{\wedge}

\newcommand{\ol}{\overline}

\newcommand{\dr}{\partial}

\newcommand{\ot}{\otimes}
\newcommand{\ap}{\approx}

\newenvironment{eqalph}{\stepcounter{equation}
\setcounter{equationa}{\value{equation}}
\setcounter{equation}{0}

\begin{eqnarray}}{\end{eqnarray}\setcounter{equation}{\value{equationa}}}

\newcounter{example}
\newcounter{remark}
\newcounter{theorem}
\newcounter{proposition}
\newcounter{lemma}
\newcounter{corollary}
\newcounter{definition}

\def\thedefinition{\arabic{definition}}

\newcommand{\mar}[1]{}

\begin{document}
\hbox{}

{\parindent=0pt

{\large \bf Gauge conservation laws in higher-dimensional 
Chern--Simons models} 
\bigskip 

{\bf G.Sardanashvily}

\medskip

\begin{small}

Department of Theoretical Physics, Moscow State University, 117234
Moscow, Russia

E-mail: sard@grav.phys.msu.su

URL: http://webcenter.ru/$\sim$sardan/
\bigskip

{\bf Abstract.}
Being gauge non-invariant, a Chern--Simons $(2k-1)$-form seen as a 
Lagrangian of gauge theory on 
a $(2k-1)$-dimensional manifold leads to the gauge conservation law of a
modified Noether current.

\end{small}
}
\bigskip
\bigskip

One usually considers the Chern--Simons (henceforth CS) gauge theory on
a principal bundle over a 3-dimensional manifold whose 
Lagrangian is the local 
CS form derived from the local
transgression formula for the second Chern form. 
This Lagrangian fails to be globally defined unless the above mentioned
principal bundle is trivial. Though the local CS Lagrangian is not
gauge-invariant, it leads to a (local) conservation law of the modified
Noether current \cite{bor98,book,sard97}. This result is also extended 
to the global CS theory on a 3-dimensional manifold \cite{al,bor2}. 
Its Lagrangian is well defined, but depends on a background gauge
potential. Therefore, it is gauge-covariant.
At the same time, the corresponding Euler--Lagrange operator 
is gauge invariant, and the above mentioned gauge
conservation law takes place. We aim to show that any higher-dimensional
CS theory admits such a conservation law.

There is the following general scheme of describing Lagrangian systems
on fibre bundles whose Euler--Lagrange operators are gauge-invariant,
but Lagrangians need not be so \cite{infin} (see also \cite{bor}).  

Let $Y\to X$ be a smooth fibre bundle over an $n$-dimensional base $X$.
An $r$-order Lagrangian is defined as a density $L=\cL d^nx$
on the $r$-order jet manifold $J^rY$ of sections of $Y\to X$
(e.g., \cite{book,epr}). Let $u$
be a vertical vector field on $Y\to X$ seen as 
an infinitesimal generator of a 
local one-parameter group of vertical bundle
automorphisms (gauge transformations) of $Y\to X$. Let $J^ru$ be 
its prolongation onto $J^rY\to X$. The following three formulas are
called into play.

(i) Let $\bL_{J^ru}L$ denote the Lie derivative of $L$
along $J^ru$. The first variational formula provides its
canonical decomposition
\mar{r1}\beq
\bL_{J^ru}L=u\rfloor\dl L + d_H(J^ru\rfloor H_L), \label{r1}
\eeq
where $\dl L$ is the Euler--Lagrange operator of
$L$, $d_H$ is the total differential, $H_L$ is a
Poincar\'e--Cartan form of $L$, and
$\gj_u=J^ru\rfloor H_L$
is the corresponding Noether current along $u$.

(ii) The master identity
\mar{r6}\beq
\dl(\bL_{J^ru}L)=\bL_{J^{2r}u}(\dl L) \label{r6}
\eeq
shows that the Euler--Lagrange
operator $\dl L$ of $L$ is gauge-invariant 
iff the Lie derivative $\bL_{J^ru}L$ is a variationally trivial
Lagrangian.  

(iii) 
An $r$-order Lagrangian $L$ is variationally trivial iff it
takes the form
\mar{r7}\beq
L=d_H\zeta +h_0(\vf) \label{r7}
\eeq
where $\zeta$ is an $(n-1)$-form of jet order $r-1$, $\vf$ is a closed
$n$-form on $Y$, and $h_0$ is the horizontal projection.
  
In the CS theories, $Y\to X$ is an affine bundle. Therefore, 
its de Rham cohomology equals that of $X$. 
Then, one obtains from the formulas (\ref{r6}) -- (\ref{r7}) that
the Euler--Lagrange operator $\dl L$ of a Lagrangian $L$ is 
gauge-invariant iff the Lie derivative $\bL_{J^ru}L$
of this Lagrangian takes the form 
\mar{r8}\beq
\bL_{J^ru}L=d_H\si +\f, \label{r8}
\eeq
where $\f$ is a non-exact $n$-form on the base $X$. 
Let the Lie derivative (\ref{r8}) reduces to
the total differential 
\mar{r4}\beq
\bL_{J^ru}L=d_H\si, \label{r4}
\eeq
e.g., it vanishes iff a Lagrangian $L$ is gauge-invariant.
Then, the first variational formula (\ref{r1}) on Ker$\,\dl L$ leads to
the equality
\mar{r5}\beq
0\ap d_H(\gj_u-\si), \label{r5}
\eeq
regarded as a conservation law of the modified Noether current
$\ol\gj=\gj_u-\si$. 

Let us write the above formulas in the case of first order Lagrangians
\cite{noninv}. Given bundle coordinates $(x^\la,y^i)$ on a fibre bundle $Y\to
X$, its first and second jet manifolds $J^1Y$ and 
$J^2Y$ are equipped with the
adapted coordinates $(x^\la,y^i,y^i_\m)$ and $(x^\la,y^i,y^i_\m,
y^i_{\la\m})$, respectively. 
We will use the notation $\om=d^nx$ and
$\om_\la=\dr_\la\rfloor \om$.
Given a first order Lagrangian 
$L=\cL(x^\la,y^i,y^i_\la)\om$
on $J^1Y$, the corresponding second order Euler--Lagrange operator reads
\mar{305}\beq
\dl L= \dl_i\cL dy^i\w\om=(\dr_i\cL- d_\la\dr^\la_i)\cL dy^i\w\om,
\label{305} 
\eeq
where $d_\la=\dr_\la
+y^i_\la\dr_i +y^i_{\la\m}\dr_i^\m$ are the total 
derivatives, which yield the total differential 
\mar{r73}\beq
d_H\vf=dx^\la\w d_\la\vf \label{r73}
\eeq
acting on the pull-back of exterior forms on $J^1Y$ onto $J^2Y$.
The kernel
Ker$\,\dl L\subset J^2Y$ of the Euler--Lagrange operator (\ref{305})
defines the Euler--Lagrange 
equations
$\dl_i\cL=0$.
Since 
\mar{r70}\beq
d_H\circ h_0=h_0\circ d, \label{r70}
\eeq
a first order
Lagrangian $L$ is variationally trivial iff 
\mar{mos11}\beq
L=h_0(\vf), \label{mos11}
\eeq
where
$\vf$ is a closed $n$-form on $Y$ and the horizontal projection $h_0$ reads
\mar{r71}\beq
h_0(dx^\la)=dx^\la, \qquad h_0(dy^i)=y^i_\la dx^\la, \qquad h_0(dy^i_\m)=
y_{\la\m}^idx^\la. \label{r71}
\eeq
Let $u=u^i\dr_i$ be a vertical vector field 
on $Y\to X$. Its jet prolongation onto $J^1Y$ is
\mar{1.21}\beq
J^1u =u^i\dr_i + d_\la u^i \dr_i^\la. \label{1.21}
\eeq
The Lie derivative of a Lagrangian $L$ along $J^1u$ 
reads
\mar{04}\beq
\bL_{J^1u}L=d(J^1u\rfloor L) + J^1u\rfloor dL=(u^i\dr_i 
+d_\la u^i\dr^\la_i)\cL\om. \label{04}
\eeq
Accordingly, the first variational formula (\ref{r1}) takes the form
\mar{bC30'}\beq
\bL_{J^1u}L= u^i\dl_i\cL\om + d_\la \gj^\la\om, \label{bC30'}
\eeq
where the Noether current along $u$ reads
\mar{r41}\beq
\gj_u=\gj^\la\om_\la= u^i\dr_i^\la\cL\om_\la. \label{r41}
\eeq

Now, let us turn to gauge theory of principal connections on a
principal bundle $P\to X$ with a structure  
Lie group $G$. Let $J^1P$ be the first order jet manifold of $P\to X$
and 
\mar{r43}\beq
C=J^1P/G\to X \label{r43}
\eeq
the quotient of $P$ with respect to the canonical action of $G$ on
$P$ (e.g., \cite{book,book00,epr}).  
There is one-to-one correspondence between the principal connections on
$P\to X$ and the sections of the fibre bundle $C$
(\ref{r43}), called the 
connection bundle. Given an atlas $\Psi$ of $P$, the connection bundle
$C$ is provided with bundle 
coordinates $(x^\la, a^r_\m)$ such that, for any its section $A$, the
local functions $A^r_\m=a^r_\m\circ A$ are coefficients of the
familiar local connection form. From the physical viewpoint, $A$
is a gauge potential.

 The infinitesimal 
generators of local one-parameter groups of vertical 
automorphism (gauge transformations) of the principal bundle $P$
are $G$-invariant vertical vector fields on $P$. There
is one-to-one correspondence between these vector fields and the sections of
the quotient 
\mar{r44}\beq
V_GP=VP/G\to X \label{r44}
\eeq
of the vertical tangent bundle $VP$ of $P\to X$ with respect to the
canonical action of $G$ on $P$. The typical fibre of
$V_GP$ is the right Lie algebra $\cG_r$ of the Lie group $G$, acting
on this typical fibre by the adjoint representation. Given an atlas
$\Psi$ of $P$ and a basis $\{\e_r\}$ for the Lie algebra $\cG_r$, we
obtain the 
fibre bases $\{e_r\}$ for $V_GP$. If $\xi=\xi^pe_p$ and
$\eta=\eta^q e_q$ are sections of $V_GP\to X$, their bracket is
\be
[\xi,\eta]=c^r_{pq}\xi^p\eta^q e_r,
\ee
where $c^r_{pq}$ are the structure constants of $\cG_r$.
Note that the connection bundle $C$ (\ref{r43}) is an affine bundle
modelled over the vector bundle $T^*X\ot V_GP$, and elements of $C$ are
represented by local $V_GP$-valued 1-forms $a^r_\m
dx^\m\ot e_r$. 
The infinitesimal generators of gauge transformations of
the connection bundle $C\to X$ are vertical vector fields
\mar{r47}\beq
\xi_C=(\dr_\m\xi^r +c^r_{pq}a^p_\m\xi^q)\dr^\m_r. \label{r47}
\eeq

The connection bundle $C\to X$ admits the canonical $V_GP$-valued 2-form
\mar{r60}\beq
\gF=(da^r_\m\w dx^\m +\frac12 c^r_{pq}a^p_\la a^q_\m dx^\la\w dx^\m)\ot
e_r, \label{r60}
\eeq
which is the curvature of the canonical connection on the principal
bundle $C\times P\to C$ (e.g., \cite{book00}). 
Given a section $A$ of $C\to X$, the pull-back  
\mar{r47'}\ben
&& F_A=A^*\gF=\frac12 F^r_{\la\m}dx^\la\w dx^\m\ot e_r,
\label{r47'}\\ 
&& F^r_{\la\m}=\dr_\la A^r_\m-\dr_\m A^r_\la +c^r_{pq}A^p_\la A^q_\m,
\nonumber
\een
of $\gF$ onto $X$ is the strength form of a gauge potential $A$.  

Turn now to the CS forms.
Let $I_k(\e)=b_{r_1\ldots r_k}\e^{r_1}\cdots \e^{r_k}$ be a $G$-invariant
polynomial of degree $k>1$ on the Lie algebra $\cG_r$ written with
respect to its basis $\{\e_r\}$, i.e.,
\be
I_k(\e)=\op\sum_j b_{r_1\ldots r_k}\e^{r_1}\cdots c^{r_j}_{pq}e^p\cdots
\e^{r_k}=
kb_{r_1\ldots r_k}c^{r_1}_{pq}\e^p\e^{r_2}\cdots \e^{r_k}=0.
\ee
Let us associate to $I(\e)$ the gauge-invariant $2k$-form 
\mar{r61}\beq
P_{2k}(\gF)=b_{r_1\ldots r_k}\gF^{r_1}\w\cdots\w \gF^{r_k} \label{r61}
\eeq
on $C$. It is a closed form due to the equalities
\be
&& dP_{2k}(\gF)=\op\sum_j b_{r_1\ldots r_k}\gF^{r_1}\w 
\cdots\w d\gF^{r_j}\w\cdots\w \gF^{r_k}= \\
&& \qquad\op\sum_j b_{r_1\ldots r_k}\gF^{r_1}\w 
\cdots\w c^{r_j}_{pq} \gF^p \w
a^q_\m dx^\m\w \cdots\w \gF^{r_k}=\\
&& \qquad (\op\sum_j b_{r_1\ldots r_k}\gF^{r_1}\w 
\cdots\w c^{r_j}_{pq} \gF^p \w \cdots\w \gF^{r_k})\w
a^q_\m dx^\m=0.
\ee
Let $A$ be a section of $C\to X$. Then, the pull-back 
\mar{r63}\beq
P_{2k}(F_A)=A^*P_{2k}(\gF) \label{r63}
\eeq
of $P_{2k}(\gF)$ is a closed characteristic form on $X$. Recall that
the de Rham cohomology of $C$ equals that of $X$ since $C\to X$
is an affine bundle. It follows that $P_{2k}(\gF)$ and $P_{2k}(F_A)$ possess
the same  
cohomology class 
\mar{r62}\beq
[P_{2k}(\gF)]=[P_{2k}(F_A)] \label{r62}
\eeq
for any principal connection $A$. Thus, $I_k(\e)\mapsto
[P_{2k}(F_A)]\in H^*(X)$ 
is the familiar Weil homomorphism.

Let $B$ be a fixed section of the connection bundle $C\to X$.
Given the characteristic form $P_{2k}(F_B)$ (\ref{r63}) on $X$, let the same
symbol stand for its pull-back onto $C$. By virtue of the equality 
(\ref{r62}), the difference $P_{2k}(\gF)-P_{2k}(F_B)$ is an exact form on $C$.
Moreover, similarly to the well-known transgression formula on a
principal bundle $P$, one can obtain the following transgression formula
on $C$:
\mar{r64,5}\ben
&& P_{2k}(\gF)-P_{2k}(F_B)=d\gS_{2k-1}(B), \label{r64}\\
&&  \gS_{2k-1}(B)=k\op\int^1_0 \gP_{2k}(t,B)dt, \label{r65}\\
&& \gP_{2k}(t,B)=b_{r_1\ldots r_k}(a^{r_1}_{\m_1}-B^{r_1}_{\m_1})dx^{\m_1}\w
\gF^{r_2}(t,B)\w\cdots \w \gF^{r_k}(t,B),\nonumber\\
&& \gF^{r_j}(t,B)=
[d(ta^{r_j}_{\m_j} +(1-t)B^{r_j}_{\m_j})\w dx^{\m_j} +\nonumber\\
&& \qquad \frac12c^{r_j}_{pq}
(ta^p_{\la_j} +(1-t)B^p_{\la_j})(ta^q_{\m_j} +(1-t)B^q_{\m_j})dx^{\la_j}\w 
dx^{\m_j}]\ot e_r.
\nonumber
\een
Its pull-back by means of a section $A$ of $C\to X$ gives
the transgression formula
\be
P_{2k}(F_A)-P_{2k}(F_B)=d S_{2k-1}(A,B)
\ee
on $X$. For instance, if $P_{2k}(F_A)$ is the characteristic Chern $2k$-form,
then $S_{2k-1}(A,B)$ is the familiar CS $(2k-1)$-form. 
Therefore, we agree to call $\gS_{2k-1}(B)$
(\ref{r65}) the CS form on the connection bundle $C$. 
In particular, one can choose the local section $B=0$. 
Then, $\gS_{2k-1}=\gS_{2k-1}(0)$ is
the local CS form. Let $S_{2k-1}(A)$ denote its pull-back 
onto $X$ by means of a section $A$ of $C\to X$. Then, the CS form
$\gS_{2k-1}(B)$ admits the decomposition
\mar{r75}\beq
\gS_{2k-1}(B)=\gS_{2k-1} -S_{2k-1}(B) +dK_{2k-1}(B). \label{r75}
\eeq

Let $J^1C$ be the first order jet manifold of the
connection bundle $C\to X$ equipped with the adapted coordinates
$(x^\la, a^r_\m, a^r_{\la\m})$. 
Let us consider the pull-back of the CS form (\ref{r65}) onto $J^1C$
denoted by the same symbol $\gS_{2k-1}(B)$, and let 
\mar{r74}\beq
\cS_{2k-1}(B)=h_0 \gS_{2k-1}(B) \label{r74}
\eeq
be its horizontal projection.
This is given by the formula
\be
&& \cS_{2k-1}(B)=k\op\int^1_0 \cP_{2k}(t,B)dt, \\
&& \cP_{2k}(t,B)=b_{r_1\ldots r_k}(a^{r_1}_{\m_1}-B^{r_1}_{\m_1})dx^{\m_1}\w
\cF^{r_2}(t,B)\w\cdots \w \cF^{r_k}(t,B),\\
&& \gF^{r_j}(t,B)= \frac12[
ta^{r_j}_{\la_j\m_j} +(1-t)\dr_{\la_j}B^{r_j}_{\m_j}
- ta^{r_j}_{\m_j\la_j} -(1-t)\dr_{\m_j}B^{r_j}_{\la_j})+\\
&& \qquad \frac12c^{r_j}_{pq}
(ta^p_{\la_j} +(1-t)B^p_{\la_j})(ta^q_{\m_j} +(1-t)B^q_{\m_j}]dx^{\la_j}\w 
dx^{\m_j}\ot e_r.
\ee

Now, let us consider the CS gauge model on a $(2k-1)$-dimensional base
manifold $X$ whose Lagrangian 
\mar{r80}\beq
L_{\rm CS}=\cS_{2k-1}(B) \label{r80}
\eeq
is the CS form (\ref{r74}) on $J^1C$. Clearly, this Lagrangian is not
gauge-invariant. 
Let $\xi_C$ (\ref{r47}) be the infinitesimal generator of gauge
transformations of the connection bundle $C$. 
Its jet prolongation onto $J^1C$ is
\be
J^1\xi_C=\xi^r_\m\dr^\m_r + d_\la\xi^r_\m\dr^{\la\m}_r.
\ee
 The Lie derivative of the Lagrangian
$L_{\rm CS}$ along $J^1\xi_C$ reads
\mar{r81}\beq
\bL_{J^1\xi_C} \cS_{2k-1}(B)= J^1\xi_C\rfloor d(h_0\gS_{2k-1}(B))
=\bL_{J^1\xi_C}(h_0 \gS_{2k-1}(B)). 
\label{r81}
\eeq
A direct computation shows that
\be
\bL_{J^1\xi_C}(h_0 \gS_{2k-1}(B))=h_0(\bL_{\xi_C}\gS_{2k-1}(B))
=h_0(\xi_C\rfloor d\gS_{2k-1}(B)+ d(\xi_C\rfloor\gS_{2k-1}(B)) .
\ee
By virtue of the transgression formula (\ref{r64}), we have
\be
d(\xi_C\rfloor d\gS_{2k-1}(B))=\bL_{\xi_C}(d\gS_{2k-1}(B))=
\bL_{\xi_C}P_{2k}(\gF)=0.
\ee
It follows that $\xi_C\rfloor d\gS_{2k-1}(B)$ is a closed form on $C$, 
i.e.,
\be
\xi_C\rfloor d\gS_{2k-1}(B)=d\psi + \vf,
\ee
where $\vf$ is a non-exact $(2k-1)$-form on $X$. Moreover, $\vf=0$ since
$P(\gF)$, $k>1$, does not contain terms linear in $da^r_\m$.
Hence, the Lie derivative (\ref{r81}) takes the form (\ref{r4}) where
\be
\bL_{J^1\xi_C} \cS_{2k-1}(B)=d_H \si,
\qquad \si=h_0(\psi + \xi_C\rfloor\gS_{2k-1}(B)).
\ee
As a consequence, the CS theory with the Lagrangian (\ref{r80}) admits
the conservation law (\ref{r5}). 

In a more general setting, one can
consider the sum of the CS Lagrangian (\ref{r80}) and some
gauge-invariant Lagrangian.

For instance, let $G$ be a semi-simple group and $a^G$ 
the Killing form on $\cG_r$. Let 
\mar{r48}\beq
P(\gF)=\frac{h}{2}a^G_{mn}\gF^m\w \gF^n \label{r48}
\eeq
be the second Chern form up to a constant multiple. Given a section $B$ of
$C\to X$, the transgression formula
(\ref{r64}) on $C$ reads
\mar{r49}\beq
P(\gF)-P(F_B)=d\gS_3(B), \label{r49}
\eeq 
where $\gS_3(B)$ is the CS 3-form up to a constant multiple. 
Let us a consider the gauge model on a 3-dimensional base manifold
whose Lagrangian is the sum 
\mar{r89}\beq
L= L_{\rm CS}+L_{\rm inv}\label{r89}
\eeq
of the CS Lagrangian
\mar{r50}\ben
&&L_{\rm CS}=h_0(\gS_3(B))=
ha^G_{mn} \ve^{\al\bt\g}[\frac12a^m_\al(\cF^n_{\bt\g} -\frac13
c^n_{pq}a^p_\bt a^q_\g)-  \label{r50}\\
&& \qquad \frac12B^m_\al(F(B)^n_{\bt\g} -\frac13
c^n_{pq}B^p_\bt B^q_\g)
-d_\al(a^m_\bt B^n_\g)]d^3x, \nonumber\\
&& \cF=h_0\gF=\frac12 \cF^r_{\la\m}dx^\la\w dx^\m\ot e_r, \qquad
\cF^r_{\la\m}=a^r_{\la\m}-a^r_{\m\la} +c^r_{pq}a^p_\la a^q_\m.
\nonumber
\een
and some gauge-invariant Lagrangian 
\mar{r85}\beq
L_{\rm inv}= \cL_{\rm
inv}(x^\la,a^r_\m,a^r_{\la\m},z^A,z^A_\la)d^3x \label{r85}
\eeq
of gauge potentials and matter fields. Then, the first variational
formula (\ref{r1}) on-shell takes the form
\mar{r86}\beq
\bL_{J^1\xi_C}L_{\rm CS}\ap d_H(\gj_{\rm CS} +\gj_{\rm inv}), \label{r86}
\eeq
where $\gj_{\rm CS}$ is the Noether current of the CS Lagrangian
(\ref{r50}) and $\gj_{\rm inv}$ is that of the gauge-invariant
Lagrangian (\ref{r85}). A simple calculation gives
\be
&& \bL_{J^1\xi_C}L_{\rm
CS}=-d_\al(ha^G_{mn}\ve^{\al\bt\g}(\dr_\bt\xi^m a^n_\g +
(\dr_\bt\xi^m +c^m_{pq}a^p_\bt\xi^q) B^n_\g))d^3x,\\
&& \gj_{\rm CS}^\al=ha^G_{mn}\ve^{\al\bt\g}(\dr_\bt\xi^m
+c^m_{pq}a^p_\bt\xi^q)(a^n_\g- B^n_\g). 
\ee
Substituting these expression into the weak equality (\ref{r86}), we
come to the conservation law 
\be
0\ap d_\al[ha^G_{mn}\ve^{\al\bt\g}(2\dr_\bt\xi^\m a^n_\g +
c^m_{pq}a^p_b a^n_\g\xi^q)+ \gj_{\rm inv}^\al]
\ee
of the modified Noether current
\be
\ol\gj=ha^G_{mn}\ve^{\al\bt\g}(2\dr_\bt\xi^\m a^n_\g +
c^m_{pq}a^p_b a^n_\g\xi^q)+ \gj_{\rm inv}^\al.
\ee


\begin{thebibliography}{ederf}

\bibitem{al} G.Allemandi, M.Francaviglia and M.Raiteri, Covariant
charges in Chern--Simons $AdS_3$
gravity, {\it Class. Quant. Grav.} {\bf 20} (2003) 483; {\it E-print
arXiv:} gr-qc/0211098. 

\bibitem{bor98} A.Borowiec, M.Ferraris and M.Francaviglia, Lagrangian
symmetries of Chern--Simons theories, {\it J. Phys. A} {\bf 31} (1998)
8823: {\it E-print arXiv:} hep-th/9801126.

\bibitem{bor} A.Borowiec, M.Ferraris, M.Francaviglia and M.Palese,
Conservation laws for non-global Lagrangians, {\it E-print arXiv}:
math-ph/0301043. 

\bibitem{bor2} A.Borowiec, M.Ferraris and M.Francaviglia, A covariant
formalism for Chern--Simons gravity, {\it J. Phys. A} {\bf 36} (2003) 2589;
{\it E-print arXiv}: math-ph/0301146.

\bibitem{book} G.Giachetta, L.Mangiarotti and G.Sardanashvily, {\it
New Lagrangian and Hamiltonian Methods in Field Theory} (World Scientific,
Singapore, 1997).

\bibitem{book00} L.Mangiarotti and G.Sardanashvily, {\it
Connections in Classical and Quantum Field Theory} (World Scientific,
Singapore, 2000).

\bibitem{sard97} G.Sardanashvily, Stress-energy-momentum tensor in
constraint field theories, {\it J. Math. Phys.} {\bf 38} (1997) 847.

\bibitem{epr} G.Sardanashvily, Ten lectures on jet manifold in
classical and quantum field theory, {\it E-print arXiv}: math-ph/0203040.

\bibitem{noninv} G.Sardanashvily, Noether conservation laws issue from
the gauge invariance of an Euler--Lagrange operator, but not a
Lagrangian, {\it E-print arXiv:} math-ph/0302012

\bibitem{infin} G.Sardanashvily, Noether conservation laws in infinite
order Lagrangian formalism, {\it E-print arXiv:} math-ph/0302063.

\end{thebibliography}
\end{document}